\documentclass[aip, jcp, showpacs, superscriptaddress, preprint ,11pt]{revtex4-1}
\usepackage{dcolumn,amsmath}
\usepackage{graphicx}
\usepackage{multirow} 
\usepackage{wrapfig}
\usepackage{longtable}
\usepackage{amsmath}
\usepackage{bm}
\usepackage{comment}
\usepackage[export]{adjustbox}
\begin{document}

\title{The near-infrared spectrum of ethynyl radical}

\author{Anh\ T.\ Le}\email{anhle@bnl.gov}
\author{Gregory\ E.\ Hall}\email{gehall@bnl.gov}
\affiliation{Division of Chemistry, Department of Energy and Photon Sciences, Brookhaven National Laboratory, Upton, NY 11973-5000, USA}
\author{Trevor\ J.\ Sears}\email{sears@bnl.gov, trevor.sears@stonybrook.edu}
\affiliation{Division of Chemistry, Department of Energy and Photon Sciences, Brookhaven National Laboratory, Upton, NY 11973-5000, USA}
\affiliation{Chemistry Department, Stony Brook University, Stony Brook, NY 11794, USA}

\begin{abstract}

Transient diode laser absorption spectroscopy has been used to measure three strong vibronic bands in the near infrared spectrum of the C$_2$H, ethynyl, radical not previously observed in the gas phase.  The radical was produced by ultraviolet excimer laser photolysis of either acetylene or (1,1,1)-trifluoropropyne in a slowly flowing sample of the precursor diluted in inert gas, and the spectral resolution was Doppler-limited.  The character of the upper states was determined from the rotational and fine structure in the observed spectra and assigned by measurement of ground state rotational combination differences. The upper states include a $^2\Sigma ^+$ state at 6696 cm$^{-1}$, a second  $^2\Sigma ^+$ state at 7088 cm$^{-1}$, and a $^2\Pi$ state at 7110 cm$^{-1}$. By comparison with published calculations (R. Tarroni and S. Carter, \textit{J. Chem. Phys} \textbf{119}, 12878 (2003) and \textit{Mol. Phys}. \textbf{102}, 2167 (2004)), the vibronic character of these levels was also assigned.  The observed states contain both $X^2\Sigma^+$ and $A^2\Pi$ electronic character.  Several local rotational level perturbations were observed in the excited states.  Kinetic measurements of the time-evolution of the ground state populations following collisional relaxation and reactive loss of the radicals formed in a hot, non-thermal, population distribution were made using some of the strong rotational lines observed.  The case of  C$ _{2} $H may be a good place to investigate the behavior at intermediate pressures of inert colliders, where the competition between relaxation and reaction can be tuned and observed to compare with master equation models, rather than deliberately suppressed to measure thermal rate constants.     

\end{abstract}

\date{\today}

\maketitle

\section{Introduction}
 The C$_2$H radical (ethynyl) is an intermediate of significant importance in the chemistry of hydrocarbon combustion, particularly in acetylene-air mixtures.\cite{Eiteneer2003, Boullart1996}  It is also related to the substituted ethynyl species which are the crucial intermediates in the hydrogen abstraction-acetylene addition (HACA) mechanism that leads to the formation of polyaromatic ring species on the way to soot in fuel rich environments. \cite{Parker2014, Kislov2013, Liu2015}  Recently, a HACA-like mechanism has also been invoked to explain the growth of single-walled carbon nanotubes in low temperature catalyzed-chemical vapor deposition syntheses using acetylene and inert gas flowing over mildly heated metal-containing catalysts. In this model, the adsorbed C$_2$H intermediate acts as a continuously regenerated ``catalyst'' in the process.\cite{Wang2014, Vasenkov2009, Page2015}  
 
 The radical was one of the earliest molecules identified in interstellar space via its microwave spectrum. \cite{Tucker1974} Models of interstellar chemistry\cite{Bettens1997, Parker2012a} invoke reaction pathways involving C$_2$H to explain the presence of polyaromatic hydrocarbons (PAHs) in the interstellar and star-formation regions in space.  Finally, investigations of the atmosphere of Titan, a moon of Saturn, have shown that photochemistry in its upper atmosphere leads to significant concentrations of polyynes, assumed to be produced in similar reaction sequences to those involved in the terrestrial formation of PAHs and soot, and including ethynyl-like intermediates in a chain reaction synthetic pathway. \cite{Shindo2001, Vuitton2001, Vuitton2001b}  In support of the extensive reaction network modeling in multiple environments,  rate constants have been measured for reactions of ethynyl radicals with various partners over a range of temperatures. \cite{Laufer2004,  Stephens1987, Lander1990,Shin1991, Farhat1993, Opansky1993, Peeters1995, Peeters1996, Opansky1996, Nizamov2004, Feng2013}  
 
 Following the astronomical observation of the ethynyl radical,\cite{Tucker1974} the gas-phase microwave spectrum was recorded and the molecular structure determined in the 1980's.\cite{Sastry1981, Gottlieb1983, Woodward1987}  Prior to this, laboratory observations had been limited to work in low temperature matrices.\cite{Jacox1975, Cochran1964}  The infrared spectrum of the radical is complicated by the presence of a low-lying \textit{A} $^2\Pi$ excited electronic state which splits due to the Renner-Teller effect. The three resulting electronic surfaces are vibronically coupled, so the vibrationally excited levels of the ground \textit{X} $^2\Sigma^+$ state all contain some $A$-state character.  The Curl\cite{Carrick1982, Pfeiffer1982, Carrick1983, Curl1985, Yan1987, Yan1987b, Stephens1988} group reported extensive work on the infrared spectrum in the 3$\mu$m (C-H stretching) region and Hirota\cite{Kanamori1987, Kanamori1988, Kawaguchi1988} recorded and analyzed the spectrum in the C-C stretching region and identified hot bands originating in the low frequency bending vibrational mode. All observed infrared spectra showed effects resulting from the strong vibronic interactions and local $J$-dependent perturbations due to background levels derived from combinations of lower frequency modes. Many more spectroscopic studies were reported in subsequent years and Pfelzer\textit{et al.}\cite{Pfelzer1996} survey the work up to the mid 1990's.  
 
 Detailed understanding of the vibronic interactions leading to the observed spectra has relied upon \textit{ab initio} methods to label the vibronic levels observed and to guide assignments, particularly at higher internal energies.  Peri\'{c} \textit{et al.}\cite{Peric1992} made a state-of-the-art calculation of the vibronic states up to approximately 5000 cm$^{-1}$ in 1992, and were able to identify the vibronic levels observed to that time.  Later, Tarroni and Carter\cite{Tarroni2003, Tarroni2004} reported more accurate calculations based on variational solutions of the nuclear motion on the coupled high level \textit{ab initio} potential energy surfaces. Importantly, the second of these papers reported absorption intensities to vibronically excited levels from the zero point level of the ground state thereby allowing vibronic state labels to be assigned to observed spectra with confidence in energy regions where the density of states is large. In low temperature matrices, Forney \textit{et al.}\cite{Forney1995} had reported infrared spectra to much shorter wavelengths than previously seen, and Hsu \textit{et al.}\cite{Hsu1993, Hsu1995,  Chiang1999} identified many higher lying vibronic states in laser-induced fluorescence (LIF) spectra in the $B^2A^{\prime} - X^2\Sigma ^+$ band system.   The vibronic structure out to similar energy levels was also observed in an anion photodetachment experiment by Zhou \textit{et al.}\cite{Zhou2007}   Recently, Sharp-Williams \textit{et al.} observed new gas phase spectra of C$_2$H between 3600 and 4100 cm$^{-1}$ in a jet cooled sample,\cite{Sharp-Williams2011a, Sharp-Williams2011b} assigning the spectra by comparison to the Tarroni and Carter's results\cite{Tarroni2004} and identifying multiple local perturbations in the rotational structure of the upper vibronic levels. Additional bands of C$_2$H in the 4250$-$4550 cm$^{-1}$ region have very recently been analyzed in a Fourier transform emission spectrum, and a band near 6340 cm$^{-1}$  has been characterized by diode laser spectroscopy in a concentration modulated discharge source.\cite{Tokaryk2015}
 
 Both the calculations\cite{Tarroni2004} and matrix spectrum\cite{Forney1995} show that the most intense bands in the entire infrared and near-infrared spectra of the radical lie around 7100 cm$^{-1}$ or 1.4 $\mu$m.  The vibronic levels at this energy are strongly mixed, but derive much of their intensity from the near-vertical transition to the $A$ (0,0,2) state, in a region of spectrum where no gas phase spectra have yet been reported.  We expect that the spectra in this wavelength region will be advantageous for kinetic or non-intrusive remote sensing applications.  In addition to the strength of these near-infrared transitions, efficient spectroscopic sources and detectors are available at these wavelengths, and optical fibers are well suited for coupling probe radiation in and out of the sample. In this work, we report measurements of the C$_2$H spectrum between approximately 6630 and 7135 cm$^{-1}$ by diode laser transient absorption spectroscopy.  We have identified and rotationally assigned three band systems in this region, based on ground state combination differences \cite{Gottlieb1983} and comparison to the predictions of Tarroni and Carter.\cite{Tarroni2004}
 
 In the course of our spectroscopic studies, we were initially puzzled to find that the decay rate of the vibronic ground state C$_2$H radical signals was several times slower than expected, based on published room temperature rate constants \cite{Stephens1987, Lander1990} for the reaction of  C$_2$H with our chosen photolytic precursor, CF$ _{3} $C$ _{2} $H, at the sample  pressures we had chosen for convenient collection of radical spectra. A time-dependent spectral simplification observed in Ar:CF$ _{3} $C$ _{2} $H mixtures that we attributed to vibrational thermalization was not observed in  Ar:C$_2$H$_2$ mixtures, where a dense set of both assigned and unassigned lines all decayed at comparable rates.   The ground state radical lifetimes were also longer than anticipated in acetylene, based on published thermal rate constants and the sample composition. To investigate these effects further, we modified the detection system slightly for fixed frequency kinetic studies, and undertook some pressure and dilution studies, to see the effects of competitive reaction and relaxation on the time-dependent spectra. The kinetic observations and discussions are reported following the spectroscopic observations and analysis below.
 
One important reason to characterize these strong bands of C$_2$H in the 7000 cm$^{-1}$ region is for future use in kinetic investigations.  In this preliminary study, the complex interplay of hot formation, vibronic relaxation and energy-dependent reactivity is evident but far from fully characterized.  Useful comparisons can be made with previous spectroscopic work where conditions reported for observing spectra of different hot or cold-band transitions in C$_2$H had been optimized. Additional comparisons can be made with previous kinetic spectroscopy studies of C$_2$H reactions, where either complete or negligible vibronic relaxation prior to reaction is a recognized precondition for the measurement of reliable thermal rate constants.
  
Hirota and co-workers \cite{Kanamori1987} had used 193 nm photolysis of a 2:3 mixture of C$_2$H$_2$:Ar at 2 Torr total pressure to search for C$_2$H transitions in the 1850 cm$^{-1}$ region.  They reported many lines, but none that could be assigned to cold bands of C$_2$H, unless they replaced the Ar with H$_2$, D$_2$, CO or CH$_4$ as a buffer gas.  He and N$_2$ were also reported to be ineffective at vibrational relaxation.  These observations are hard to understand given current understanding of the kinetics.  The acetylene pressure was high enough to make the estimated lifetime of C$_2$H shorter than the microsecond detector response, and adding hydrogen would seemingly only decrease the lifetime by means of an additional reactive loss channel.\cite{ Shokoohi1986, Farhat1993, Tasaki1994}  Clear spectroscopic evidence of the $\nu_3$ fundamental of C$_2$H was nevertheless obtained.

Hsu et al.\cite{Hsu1993,Hsu1995}  used pump-probe delay time-dependent LIF intensities to help assign a selection of bands originating in various vibrationally excited  states of C$_2$H ($X ^{2}\Sigma^{+}$).    They reported successively longer lifetimes for lower energy states, but no resolved growth kinetics for any bands, following 193 nm excitation of C$_2$H$_2$/H$_2$ mixtures.   In a subsequent report,\cite{Chiang1999}  the substitution of SF$_6$ for Ar, He, or H$_2$ in the C$_2$H$_2$ photolysis sample was found to increase the relative intensity of otherwise weak LIF bands originating in low bending levels of C$_2$H, consistent with enhanced vibrational relaxation by SF$_6$.  No LIF transitions originating from the ground vibrational state of C$_2$H ($X ^{2}\Sigma^{+}$) have been reported, presumably due to poor Franck-Condon factors.
 
For kinetic spectroscopy studies, vibronic relaxation prior to reaction is typically tested by varying the pressure or choice of an inert buffer to verify insensitivity of  the observed reaction rates.\cite{Stephens1987,Lander1990,Farhat1993,Opansky1993,Opansky1996}  Direct measurements of hot band relaxation\cite{Yan1987} are preferable, but not always feasible, to confirm the timescale of vibrational thermalization.
 
Using the newly assigned bands, we report the results of our investigation into pressure-dependent kinetics of state-resolved relaxation in competition reaction with two C$_2$H precursors:  C$_2$H$_2$ and CF$_3$C$_2$H. The vibronic relaxation appears to be slower than has often been claimed.

\section{Experimental Details}
 The transient absorption spectra were recorded using a spectrometer very similar to that described previously. \cite{Chang2011} Ethynyl radicals were formed by 193 nm ArF excimer laser photolysis of a sample of CF$_{3}$C$_{2}$H (trifluoromethyl acetylene, or 3,3,3 trifluoropropyne) obtained from Synquest Laboratories, diluted in argon.  A Sacher Laserteknik tunable diode laser, model TEC 500, with output centered at 1450 nm was the spectroscopic source.  One portion of the laser power was sent to a wavemeter (Burleigh WA-1500), and another portion was used as a reference beam for active noise reduction.  The main beam was directed through a Herriott-type absorption cell with 35 passes between mirrors separated by 2 meters.  Signal and reference beams were imaged onto an \textit{a.c.}-coupled dual InGaAs photoreceiver (New Focus 1617).  The excimer laser beam, with typical pulse energies of 30 mJ in a 3 cm$^{2}$ beam, passed through the center of the absorption cell and overlapped the probe beam in the central portion of the cell. We estimate the effective absorption path length through the photolysis volume was about 25 meters.  Mixtures of Ar and CF$_{3}$C$_{2}$H in a 1:1 ratio were prepared in a large storage flask at pressures of approximately one atmosphere.  This gas mixture flowed slowly at 5 sccm though the cell. Total pressures used for the spectral scans were typically 1 Torr.  Mixtures of acetylene in Ar were also used for comparison in some spectral regions, producing similar but weaker spectra with generally faster decay rates.  Because of the stronger and longer-lasting signals, the systematic spectroscopy scans were carried out with the  CF$_3$C$_{2}$H precursor.  
 
 Some additional exploratory measurements of the relaxation and reaction kinetics were made at fixed probe frequencies on selected  ground state ethynyl radical transitions, as a function of precursor and inert buffer gas (argon) concentrations. These measurements used a \textit{d.c.}-coupled balanced photoreceiver (New Focus 1817), with sequential measurement and off-line subtraction of on- and off-resonant signals to remove the reproducible background  associated with acoustic transients and increase the fidelity of the transient signals.  Tests with varying flow rates or a reduction of the excimer repetition rate below its typical value of 10 Hz showed no sign of secondary photolysis or sample depletion.

\section{Results and Analysis} 
\subsection{Spectroscopy}
\subsubsection{Results}
Initial spectroscopic searches for the strong bands observed in the low temperature matrix spectrum \cite{Forney1995} and predicted by Tarroni and Carter\cite{Tarroni2004} were carried out by photolysis of a slowly flowing sample of CF$_3$C$_{2}$H in Ar at 193 nm.  Transient absorption bands presumed to be due to C$_{2}$H were detected throughout the search region.  Further support for the attribution of absorption lines to  C$_{2}$H was obtained by observing  some of the same lines in spectra using  C$_{2}$H$_{2}$ as the photolytic precursor.    Spectra recorded immediately following the photolysis pulse were much more complicated than spectra at longer delay times, which we attribute to collisional cooling of an initially formed hot distribution.  The time development of signals in samples containing   C$ _{2} $H$ _{2} $ was quite different to that observed with CF$_3$C$_{2}$H, as might be expected due to differences in absorption cross section, quantum yields for C$ _{2} $H, the initial state distributions, relaxation kinetics, and rates of reaction with the precursors. Some further kinetic investigations are presented below, but for the purposes of the spectroscopic studies, CF$_3$C$_{2}$H was found to provide larger signals and simpler spectra after partial relaxation, when compared to C$ _{2}$H$ _{2} $ samples.  Therefore, all of the subsequent spectroscopic measurements were made using CF$_3$C$_{2}$H precursor in Ar. 

\begin{figure}[Hb]

	\includegraphics[scale=0.7,center]{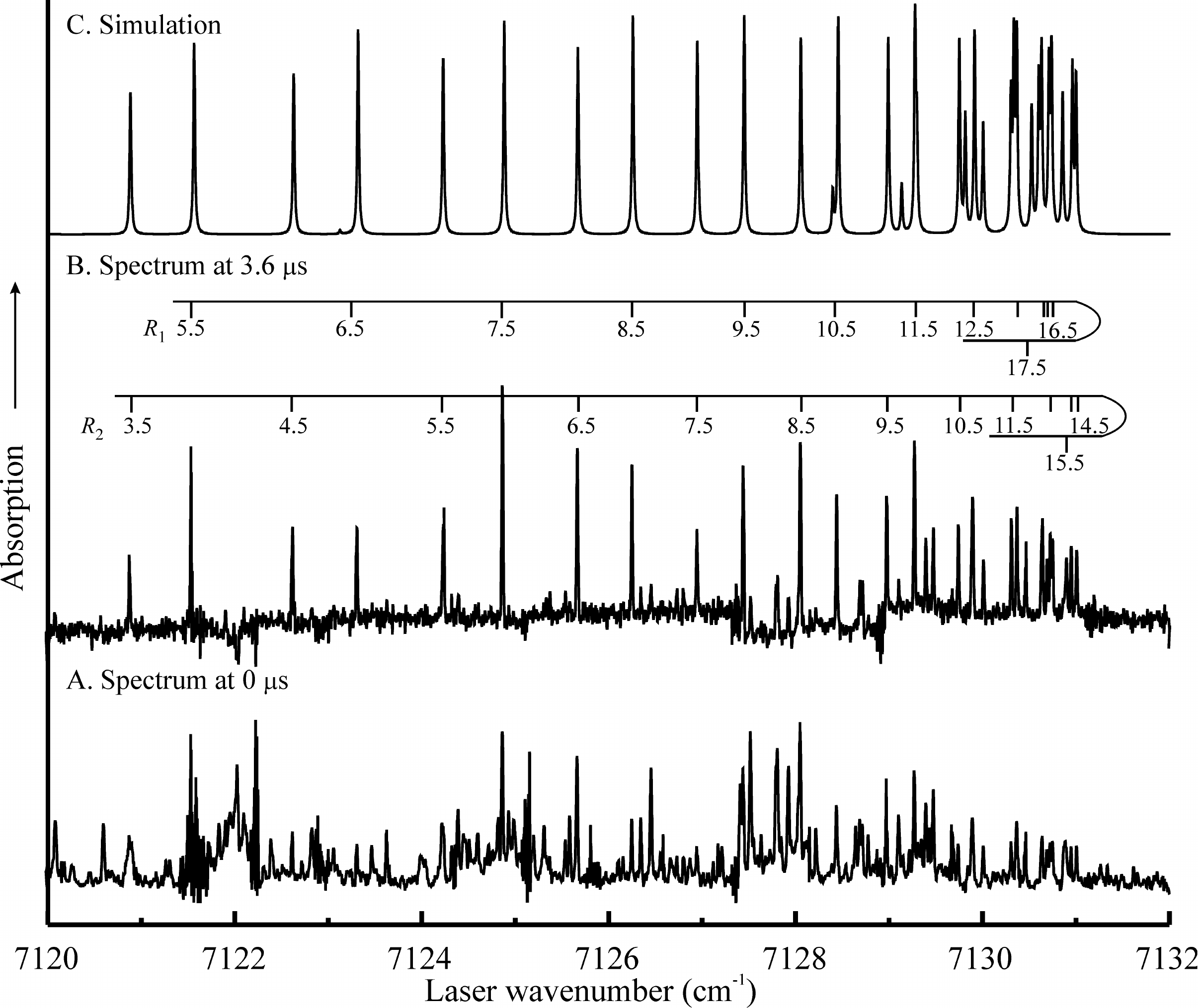}
	\centering
	\caption{\label{Pi} \footnotesize{A  15 cm$^{-1}$ portion of the time dependent transient absorption spectra of C$_{2}$H following photolysis of CF$_3$C$_2$H precursor in argon showing the  $R_{1}$ and  $R_{2}$ branches of the $^2\Pi \leftarrow\,^2\Sigma ^{+}$(000) band at 7110 cm$^{-1}$. The simulated spectrum was obtained using the optimized set of parameters in Table 4.}}
	\label{figure1.}

\end{figure}

Examples of a section of the observed spectra near 7125 cm$^{-1}$ at short and long time delays with the CF$_3$C$_{2}$H precursor are shown in Figure \ref{figure1.}. This spectrum was obtained by averaging 40 photolysis laser shots per wavelength step (typically 0.005 cm$^{-1}$) in a 1:1 mixture of CF$_3$C$_{2}$H and argon at a total pressure of 1.0 Torr. Figure \ref{figure1.}(a) shows a complex C$ _{2} $H spectrum at zero time delay following the photolysis pulse with a time gate of 1.0 $ \mu $s. After a 3.6 $ \mu $s time delay, the congestion is substantially reduced, leaving a much simpler spectrum, as shown in Figure \ref{figure1.}(b). Two branches, each with a bandhead, can be identified and assigned to  \textit{R}$ _{1} $ and \textit{R}$ _{2} $  branches of a $^2\Pi \leftarrow\, ^2\Sigma ^+$ band. Perturbations starting at \textit{R}$ _{2} $(15.5) and \textit{R}$ _{1} $(16.5) are also evident in Figure \ref{figure1.}.

\begin{figure}[H!]
	
	\includegraphics[width=0.7\linewidth, center]{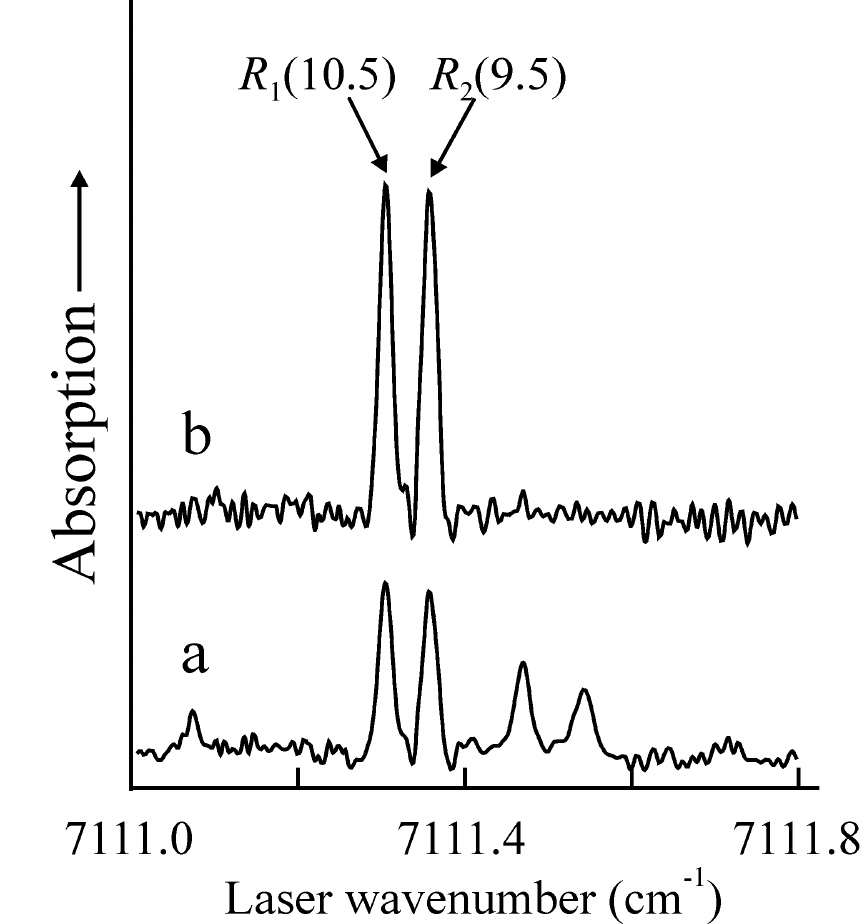}
	\centering
	\caption{\label{acetyleneprecursor} \footnotesize{Prompt and delayed spectra, illustrating growth of assigned lines and disappearance of unassigned lines. Trace (a) and (b) were obtained using 1 Torr of a 1:1 mixture of Ar:CF$_{3}$C$_{2}$H  at 0 $ \mu $s and 3.6 $\mu$s time delay, respectively.}}
	\label{figure2.}

\end{figure}

The strong band observed in the Ne matrix spectrum at 7064 cm$^{-1}$\, \cite{Forney1995} has been attributed to a combination of two bands,\cite{Tarroni2004} a $^2\Pi \leftarrow\, ^2\Sigma ^+$ band calculated at 7106 cm$^{-1}$ and a  $^2\Sigma ^+ \leftarrow\, ^2\Sigma ^+$ band at  7099 cm$^{-1}$, consistent with the sub-band identifications above.  The $^2\Sigma ^+$ lower state in both of these transitions is the ground state of the radical, and analysis of the cold spectra proceeded by searching for ground state combination differences based on the ground state constants of Gottlieb et al.\cite{Gottlieb1983} assuming vibronic bands of the type predicted.  $^2\Sigma ^+$ state energy levels are described by a rotational quantum number ($J$) and a parity label (+ or $-$) with $J = N + 0.5$ (F$_1$) or $J = N - 0.5$ (F$_2$), where $N$ is the total angular momentum excluding spin. The lowest level $N=0$ ($J=0.5$,  F$_{1}$) has (+) parity and the parity labeling alternates with $N$ for $^2\Sigma ^+$ states. As a result, \textit{ P-} and \textit{R-} branches are doubled. The doublet splitting is illustrated in Figure \ref{figure2.} (b) for lines in the $^2\Sigma ^+\leftarrow\, ^2\Sigma ^+$ transition at 7111.4 cm$^{-1}$ and is due to the difference in the spin-rotation splittings between the ground and the excited states.  The early time spectrum in Figure \ref{figure2.}(a) shows additional unassigned, presumably hot band, lines that are gone at the time of the delayed spectrum shown in Figure \ref{figure2.}(b). A more detailed comparison of the time dependence of ground state and vibronically excited states for different sample mixtures and pressures is presented in section \ref{sec:kinetics} below. 

The energy levels of the $^2\Pi$ state are also labeled by rotational ($J$) and parity quantum numbers. Two sub-states $^2\Pi_{1/2}$ and $^2\Pi_{3/2}$ result from spin-orbit coupling.  Analysis below shows the spin-orbit constant, A, is small and negative, therefore the $^2\Pi_{3/2}$ and  $^2\Pi_{1/2}$ energy levels map onto F$_1$ and F$_2$, respectively. As a result, 6 main branches (P$_1$, P$_2$, Q$_1$, Q$_2$, R$_1$, and R$_2$) were observed. The four additional satellite branches, with F$_1 \leftrightarrow \, $F$_2$ were not observed here. 

\begin{figure}[H!]
	\includegraphics[scale=0.7, center]{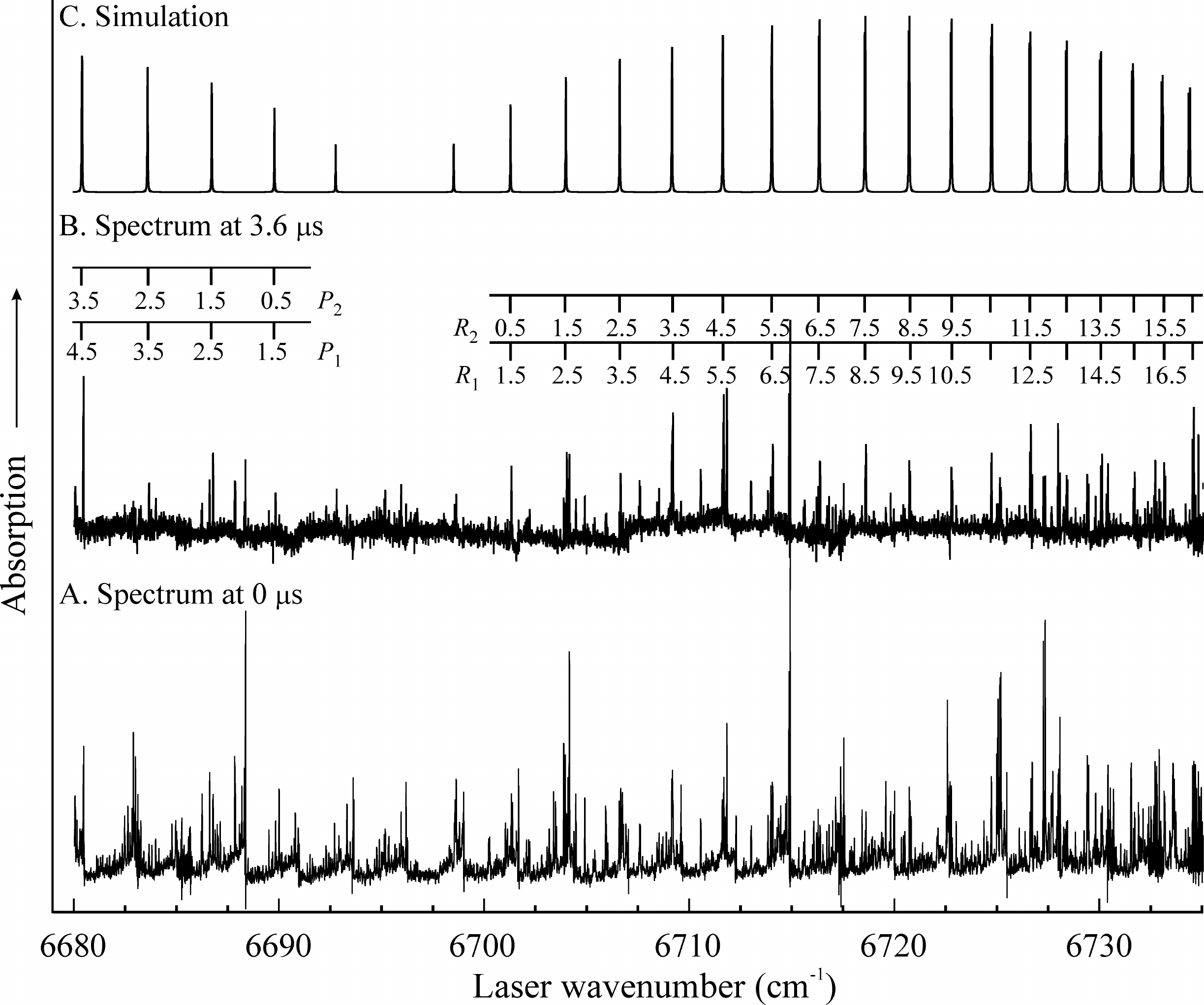}
	\centering
	\caption{\footnotesize{A 55 cm$^{-1}$ portion of the time dependent transient absorption spectra of the $^{2}\Sigma \leftarrow\,^{2}\Sigma ^{+}$(000) band of C$_{2}$H at 6696 cm$^{-1}$. The calculated spectrum was obtained using the optimized set of parameters in Table 4.}}
	\label{figure3.}
\end{figure}

Seeking further support for some of the $^2\Pi \leftarrow\, ^2\Sigma ^+$ assignments, we searched the 6700 cm$^{-1}$ region for hot band transitions coming from the \textit{X} (0,1$^{1}$,0) ($\Pi$ symmetry) to the same upper level states. The fundamental bending vibrational frequency is 371.6034(4) cm$^{-1}$; rotational and fine structure parameters for this level were determined by Kanamori et. al. \cite{Kanamori1988}  We did not find these hot band transitions above our noise level at the calculated frequencies, likely due to poor Franck-Condon factors. Instead, an additional  $^2\Sigma ^+ \leftarrow\, ^2\Sigma ^+$ band was identified in this region. Figure \ref{figure3.} shows a portion of this spectrum at 0 $\mu$s and 3.6 $\mu$s, with some \textit{R}$ _{1} $, \textit{R}$ _{2}$, \textit{P}$ _{1} $ and \textit{P}$ _{2} $ branch lines labeled.  The band origin of this $^2\Sigma ^+ \leftarrow\, ^2\Sigma ^+$ is at 6695.678 cm$ ^{-1}$. Complete data files of the two spectral regions between 6630 and 7135 cm$^{-1}$ studied here, at three different times (0, 2.6 and 3.6 $ \mu $s) are included as supplementary material.\textbf{(supplemental material)}

\subsubsection{Analysis}
The $^2\Sigma ^+$ rotational Hamiltonian can be written as:
\begin{equation} \label{eq1}
 H =B\mathbf{\hat{N}} ^2-D\mathbf{\hat{N}}^4 +\gamma(\mathbf{\hat{N}} \cdot \mathbf{\hat{S}})
\end{equation}
leading to the energy level expression:
\begin{equation} \label{eq2}
\begin{split}
 E(F_{1}) &=BN(N+1) -DN^2(N+1)^2 +\frac{\gamma N}{2}  \\
 E(F_{2}) &=BN(N+1) -DN^2(N+1)^2 -\gamma \frac{(N+1)}{2} 
\end{split}
\end{equation} 
The B$^{\prime\prime}$, D$^{\prime\prime}$, and $ \gamma^{\prime\prime}$  rotational and spin-rotational ground state constants were fixed to the literature values \cite{Gottlieb1983}.

The energies for the $^2\Pi$ state were modeled by including  the spin-orbit interaction (A), rotational constant (B) quartic centrifugal distortion correction (D), the spin-rotation ($ \gamma $), and the $ \Lambda $-doubling ($p+2q$). In the Hund's case (a) basis limit, the effective Hamiltonian for the  $^2\Pi$ state is \cite{Brown2003}:

\begin{equation}
\label{eq3}
\begin{split}
H _{eff} (^2\Pi)  = A(L_{z}\cdot S_{z})  + B\mathbf{\hat{R}} ^2 + D(\mathbf{\hat{R}}^2)^2
+\gamma(\mathbf{(\hat{J}- \hat{S})} \cdot \mathbf{\hat{S}} ) \\ 
+ \frac{1}{2} (p+2q)(e^{-2i\phi}\, \mathbf{\hat{J}} _{+}  \mathbf{\hat{S}}_{+} + e^{+2i\phi} \, \mathbf{\hat{J}}_{-} \mathbf{\hat{S}} _{-} )\\
\mathrm{where} \, \, \mathbf{\hat{R}}=(( \mathbf{\hat{J}} - \mathbf{\hat{S}}) - \mathbf{\hat{L}})
\end{split}
\end{equation}

In equation \ref{eq3}, $\mathbf{\hat{J}}_{\pm}$ and $\mathbf{\hat{S}}_{\pm}$ are the shift operators of $\mathbf{\hat{J}}$, the total angular momentum in the absence of nuclear spin, and of $\mathbf{\hat{S}}$, the total electron spin, and  $\phi$ is the azimuthal coordinate of the unpaired electron. The energies of the $^2\Pi$ state were determined by numerical diagonalization of the Hamiltonian matrix representation of dimension 4x4 constructed in a sequentially coupled, non-parity conserving Hund's case (a) basis set. The energy levels can be modeled using either the Hund's case (a) basis set, which is a good approximation for the low-rotational levels or  Hund's case (b) basic set, which is better for higher rotational levels (N $\geq $ 11). The calculated energy levels are the same in both Hund's case (a) and case (b) limits.\, \cite{Sharp-Williams2011b}

In the $^2\Sigma ^+ \leftarrow\, ^2\Sigma ^+$ band  at 7088 cm$^{-1}$, 80 transition wavenumbers  were identified via combinational differences using the ground state parameters constants of Gottlieb et al.\cite{Gottlieb1983}  However, only 32 transition wavenumbers were entered directly into the fit to the upper state parameters, due to multiple perturbations in the upper state energy levels.  Deviations in excess of 0.015 cm$^{-1}$, more than three times the measurement error, were considered evidence of local perturbations strong enough to exclude a measured transition from the fit. The measured line centers are presented in Table \ref{table1}, along with the assignments and differences between the calculated and observed transition wavenumbers. Similarly, 66 transition wavenumbers in the $^2\Sigma ^+ \leftarrow\, ^2\Sigma ^+$ band  at 6696 cm$^{-1}$ were identified by combination differences using the same ground state parameters, and 49 transition wavenumbers were used in the fit. The line centers are presented in Table \ref{table2}, along with the assignments and differences between the calculated and observed transition wavenumbers. The transitions from $^2\Sigma ^+\leftarrow\, ^2\Sigma ^+$ bands at 6696 and  7088 cm$^{-1}$ were least squares fit to the simple energy expressions in Equation \ref{eq2}, described above, to determine the excited state constants including B$^{\prime}$, D$^{\prime}$, $ \gamma ^{\prime}$ and the origin for both upper $ ^{2}\Sigma ^{+}$ states. The standard deviation of the fits were 0.004 cm$ ^{-1} $ and 0.009 cm$ ^{-1} $  for the bands at  6696 and 7088 cm$^{-1}$  respectively. The optimized parameters and associated errors are given in Table \ref{table4}.

In the $^2\Pi \leftarrow\, ^2\Sigma ^+$ band at 7110 cm$^{-1}$, 69 transition wavenumbers  were identified via combinational differences using the ground state parameters. Of these, 57 were included in the least squares fit to determine the excited state parameters \textit{A}$^{\prime}$, \textit{B}$^{\prime}$, \textit{D}$^{\prime}$, $\gamma\,^{\prime}$, $(p+2q)\,^{\prime}$, and the band origin. The standard deviation of the fit is 0.010 cm$ ^{-1} $. The optimized parameters and associated errors are given in Table \ref{table4}.

\subsection{Kinetics}
\label{sec:kinetics}
 
\begin{figure}[H!]
	\includegraphics[scale=0.7, center]{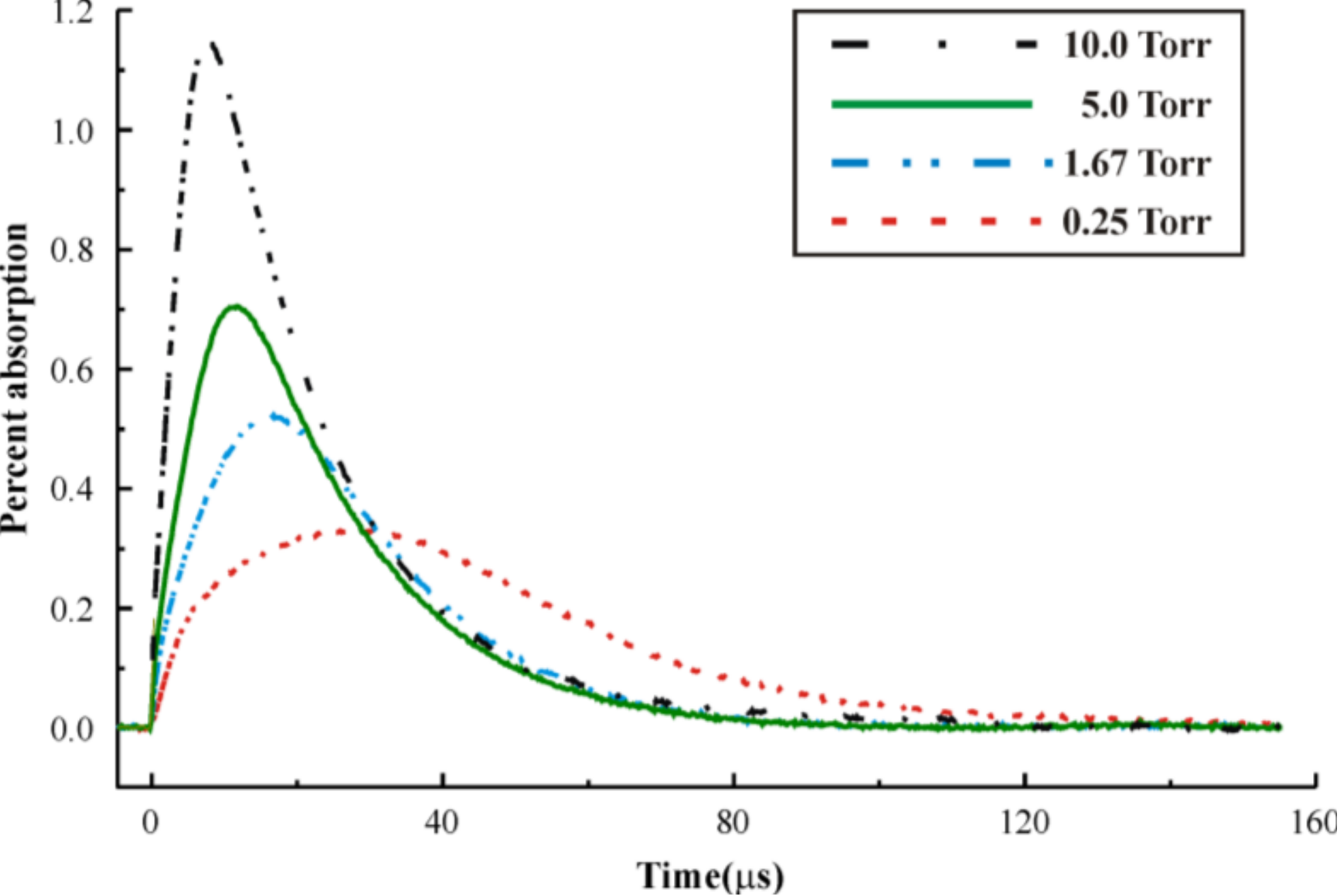}
	\centering
	\caption{\footnotesize{Transient absorption signals measured on the \textit{R}(N=4) line at 7100.303 cm$ ^{-1} $ for samples of 0.05 Torr CF$ _{3} $C$ _{2} $H diluted in Ar at the total pressures as labeled.}}
	\label{figure4.}
\end{figure}

The transient absorption of C$ _{2} $H transitions for different photolytic precursors and buffer gas pressures offers a view of the competition between collisional thermalization and reactive loss.  Figure \ref{figure4.} illustrates the time-dependent absorption signals observed on the \textit{R}(4) line of the 7088 cm$ ^{-1} $  band at a constant CF$ _{3} $C$ _{2} $H precursor partial pressure of 50 mTorr with decreasing amounts of Ar buffer gas.  Small corrections have been applied to normalize for changes in photolysis and probe laser intensity variations, so that the percent absorption signals are directly comparable. The rise rate becomes progressively slower as the Ar pressure decreases, and the peak amplitude decreases.  At 5 and 10 Torr, the decay rate is about the same, and in good agreement with the published  \cite{Stephens1987, Lander1990} room temperature rate constant of 3.5$\times$10$^{-11}$ cm$ ^{3} $ molec$ ^{-1} $ s$ ^{-1} $ for the C$ _{2} $H + CF$ _{3} $C$_{2} $H reaction, and the partial pressure of  CF$ _{3} $C$ _{2} $H.  The delayed maximum and slower decay at the lowest total pressure of 0.25 Torr superficially resembles the expectations of slower vibrational relaxation.  When the vibrational relaxation cascade is no longer rapid compared to the reactive loss, the observed decay rate will no longer be a measure of the reaction rate, as the relaxation will continue to provide a delayed source of radicals in the probed energy level.  The absolute amplitudes tell an additional story, however.   If the changing Ar pressure only affected the thermalization rate, and the reaction rate depended only on the constant  partial pressure of CF$ _{3} $C$_{2}$H,  the observed time dependence would be described by an exponentially decaying total population of C$ _{2} $H times a monotonically increasing, time-dependent fractional population in the detected state.  A signal probing C$ _{2} $H (0,0,0) at low Ar pressure could never exceed the signals at higher Ar pressure  under these conditions, contrary to our observations.  The most likely explanation appears to be that the unrelaxed population in higher energy states does not react as rapidly as the thermalized C$ _{2} $H radicals, so that delayed relaxation extends the lifetime of the total  C$ _{2} $H radical pool.

A selection of R-branch lines probing rotational states from $N$=4 to $N$=15 in the 7088 cm$^{-1}$ band were measured to assess the time-dependent rotational distributions, and to test for the possibility of high rotational states affecting the kinetics.  For a 1\% mixture of CF$_3$C$_2$H in Ar at 5 Torr, the normalized time-dependent signals were superimposable throughout the rise and fall, as one would expect for rotational thermalization faster than the $\mu$s time scale.  A plot of the data is included in the \textbf{supplementary material, Figure S1}.  We note that the apparent temperature of the stationary rotational distribution would be 270 $\pm$ 10 K if the rotational line intensities are interpreted with unperturbed line strength factors.  The apparently colder than ambient rotational temperature can be explained by an increasing rotational mixing with spectroscopically dark levels at higher $J$. The upper level of this transition is a $^2\Sigma$ state whose $A-$state character is predominantly  $A\,(030)\kappa\,^2\Sigma$.\cite{Tarroni2004}  This level will be mixed with the dark $A\,^2\Delta$ components of the same (030) level in the Renner-Teller mixed $A$ state, via a \textit{K}-type resonance\cite{Bolman1975} that includes a $J^2-$dependent mixing term.  We have modeled this interaction and confirmed that a mixing coefficient defined as a fractional contribution of dark state wavefunction of $3.5 \times 10^{-4}$, when  combined with this $J-$dependence, accounts the observed reduction in the intensity of the higher $J$ lines in the band.

\begin{figure}[h]
	\includegraphics[scale=0.7, center]{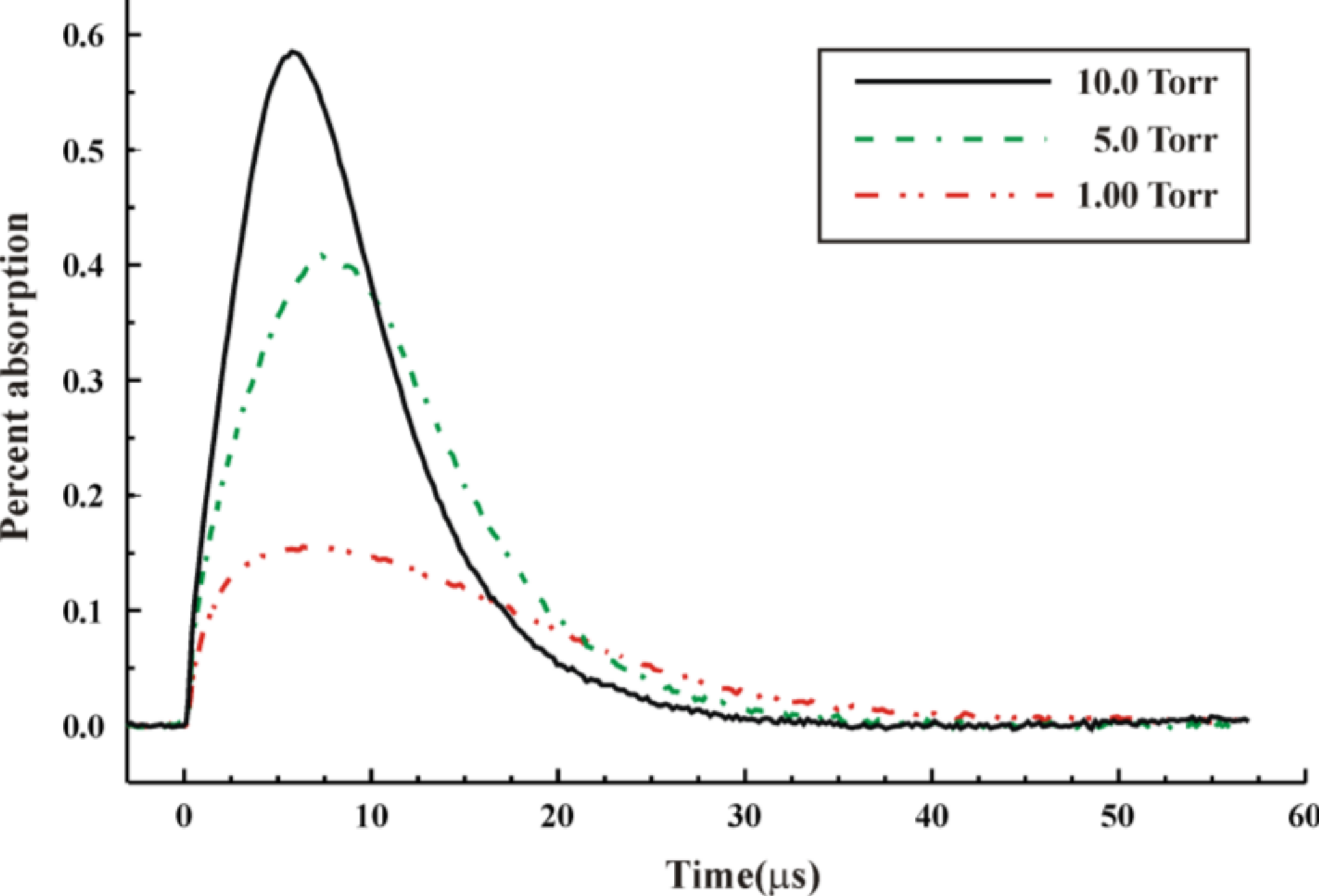}
	\centering
	\caption{\footnotesize{Transient absorption signals measured on the \textit{R}(N=4) line at 7100.303 cm$ ^{-1} $ for samples of 0.05 Torr C$ _{2} $H$ _{2} $ diluted in Ar at the total pressures as labeled.}}
	\label{figure5.}
\end{figure}

An analogous but contrasting set of pressure dependent measurements in Argon/acetylene mixtures is illustrated in Figure \ref{figure5.}. As with  CF$ _{3} $C$ _{2} $H, increasing the Ar partial pressure at fixed precursor pressure makes the signal rise more quickly to a larger maximum and a faster decay.  Higher pressures are required to reach a limiting decay rate, which at 10 Torr, is still about 30\% slower than expected, given the 50 mTorr partial pressure of C$_2$H$_2$ and a room temperature rate constant of 1.3 x 10$^{-10}$ cm$^3$ molec$^{-1}$ s$^{-1}$. $\,$\cite{Stephens1987, Shin1991, Opansky1996, Laufer2004}  Apparently even in the presence of Ar at 10 Torr the vibrational thermalization is not yet complete on the time scale of reaction with C$_2$H$_2$.

\begin{figure}[h]
	\includegraphics[scale=0.7, center]{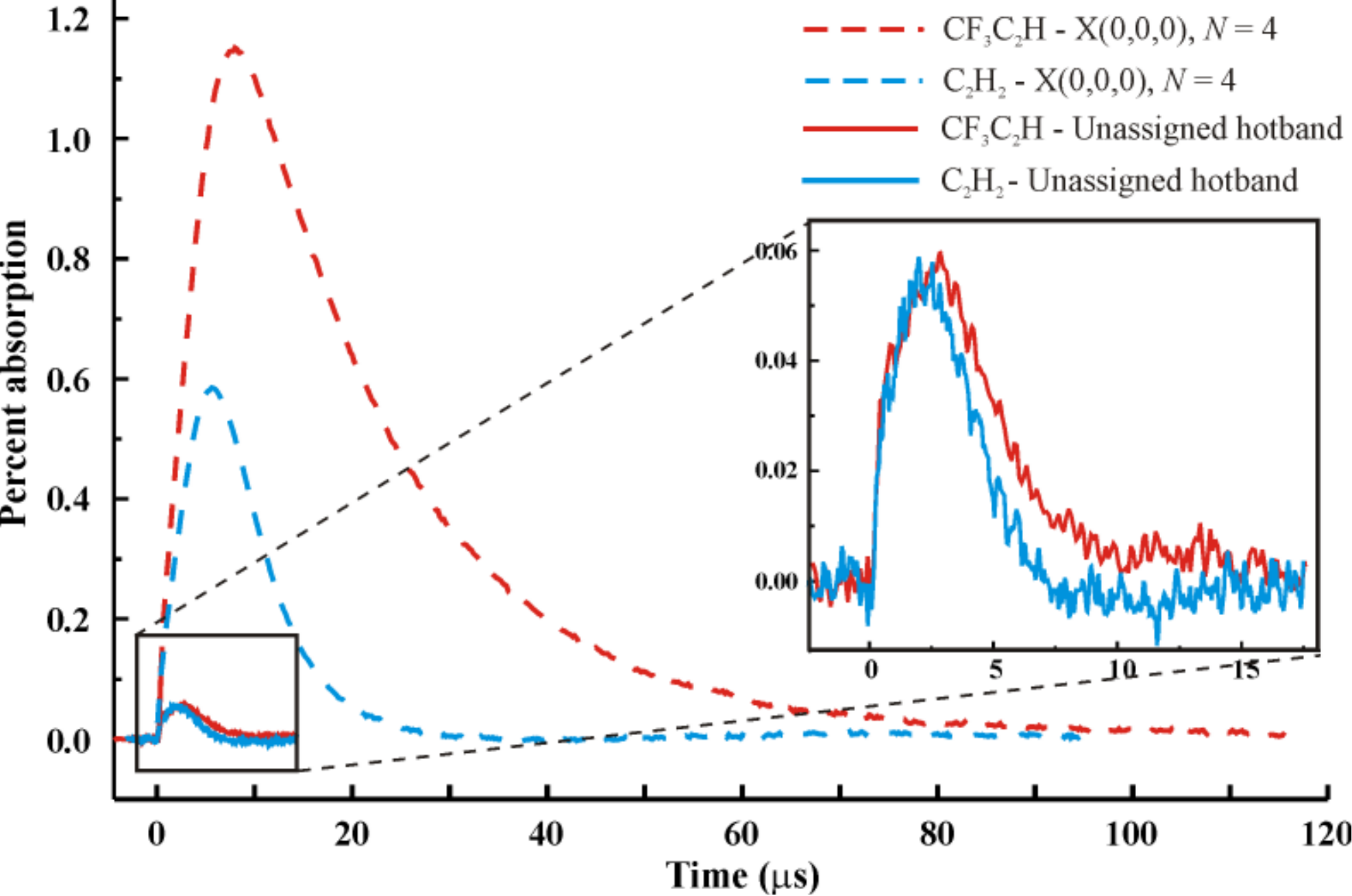}
	\centering
	\caption{\footnotesize{Comparison of C$_2$H relaxation kinetics in trifluoropropyne and acetylene, showing transient C$_2$H absorption signals for the  \textit{X}(0,0,0),\textit{N}=4 line at 7100.303 cm$ ^{-1} $ and an unassigned hot band line observed at 7111.470 cm$ ^{-1} $ using 0.05 Torr of CF$ _{3} $C$ _{2} $H or C$ _{2} $H$ _{2} $ diluted in Ar at 10 Torr total pressure.}}
	\label{figure6.}
\end{figure}

The differences between CF$ _{3} $C$ _{2} $H and C$_2$H$_2$ can be further illustrated by comparing the growth and decay of an absorption signal originating in the vibrationless state of C$_2$H with an unassigned hot band line observed at 7111.470 cm$^{-1}$ with either precursor. This line was illustrated in Figure \ref{figure2.}, and is typical of many, observed to decay quickly as the ground state lines grew in. Figure \ref{figure6.} compares these signals for two different probed transitions on samples of the two photolytic precursors, each prepared as a 200:1 mixture of Ar:precursor, measured at 10 Torr total pressure. The large signals probe the $N = 4$ level of the ground state, and illustrate the smaller signals and faster reaction for  C$_2$H$_2$  than for CF$ _{3} $C$ _{2} $H.  The smaller, faster signals, expanded in the inset figure contrast the hot band kinetics in the same two samples.  Without knowing the exact energy of the probed state, no quantitative conclusions can be drawn, but it can be seen that even a 200:1 ratio of Ar buffer to precursor is insufficient open up a clear separation of time scales for vibrational relaxation and reactive loss of the vibrationless C$_2$H in the case of  C$_2$H$_2$, since the decays of the vibrationless and hot band levels are similar, while it is adequate in the case of CF$_{3}$C$ _{2}$H.  This may be due to an initially higher extent of internal excitation in the acetylene photodissociation, perhaps also affected by differences in the intrinsic competition between reaction and relaxation for collisions of the hot radicals with the two different precursor molecules. The initial C$_2$H internal energy distribution of the photofragments produced in acetylene photodissociation has been characterized by H atom Rydberg time-of-flight measurements. \cite{Zhang1995, Mordaunt1998} The structured kinetic energy spectra indicate preferential excitation in high bending states, low rotational excitation, and  a significant population at peaks near the energy of the $A$ state.  The available energy for C$_2$H fragments in the 193 nm photodissociation of CF$_3$C$_2$H is about 30\% higher (198 kJ mol$^{-1}$ vs. 151 kJ mol $^{-1}$), \cite{Fahr2003} but the additional degrees of freedom in the CF$_3$ co-fragment compared to H atoms provide statistical grounds to expect a cooler C$_2$H internal energy distribution from CF$_3$C$_2$H than from C$_2$H$_2$.  No direct comparison has been reported, however, to our knowledge.

A striking difference in the relative intensities of prompt and delayed spectra under the high concentration conditions initially used to investigate the spectroscopy was one of the observations that led us to the current kinetic measurements.  In our preliminary spectroscopic measurements, a substantial spectral simplification was observed in the CF$_3$C$_2$H:Ar samples during the first few microseconds, as illustrated in Figures \ref{figure1.}-\ref{figure3.}.  Selected short scans of C$_2$H$_2$:Ar samples at the same 1:1 mixing ratio and total pressure of 1 Torr showed the same transitions, different initial relative intensities, a more rapid decay, but no significant differences in the decay rates of the lines assignable to ground state levels and other unassigned lines that behaved like hot bands in the  CF$_3$C$_2$H:Ar sample.   We now attribute this behavior to the dominance of the reaction of C$_2$H with C$_2$H$_2$ in competition with less efficient  vibrational relaxation by either Ar or C$_2$H$_2$. We believe most of the reactive loss seen under these conditions is occurring from vibrationally excited C$_2$H.

\section{Discussion}
\subsection{Spectroscopy}

Previously, there have been no reported gas phase measurements of C$ _{2} $H in the 6700 cm$ ^{-1}$ and 7100 cm$ ^{-1} $ regions studied here. However, the  origins determined for all three bands identified and assigned are in good agreement with the calculated values from Tarroni et al. \cite{Tarroni2004} Compared to the earlier matrix work,\cite{Forney1995} a new $^2\Sigma ^+ \leftarrow\, ^2\Sigma ^+$  band was identified and assigned to the transition from the ground state to the X(0,8$^\text{0}$,2) state again based on Tarroni and Carter.\cite{Tarroni2004} This state has a very small fraction of $A\,^2\Pi$ electronic state mixing, and was assigned to the vibrational combination level of $X$ state. The standard deviation for the fit of this band is smaller than the other two observed bands in this region, possibly reflecting its less mixed character. We estimate the overall intensity ratio between the three bands at 6696, 7088 and 7110 cm$^{-1}$ is about 1:4:2, respectively, which is in good agreement with the calculated intensity ratios.\cite{Tarroni2004}

\begin{figure}[H!]
		\includegraphics[scale=0.7, center]{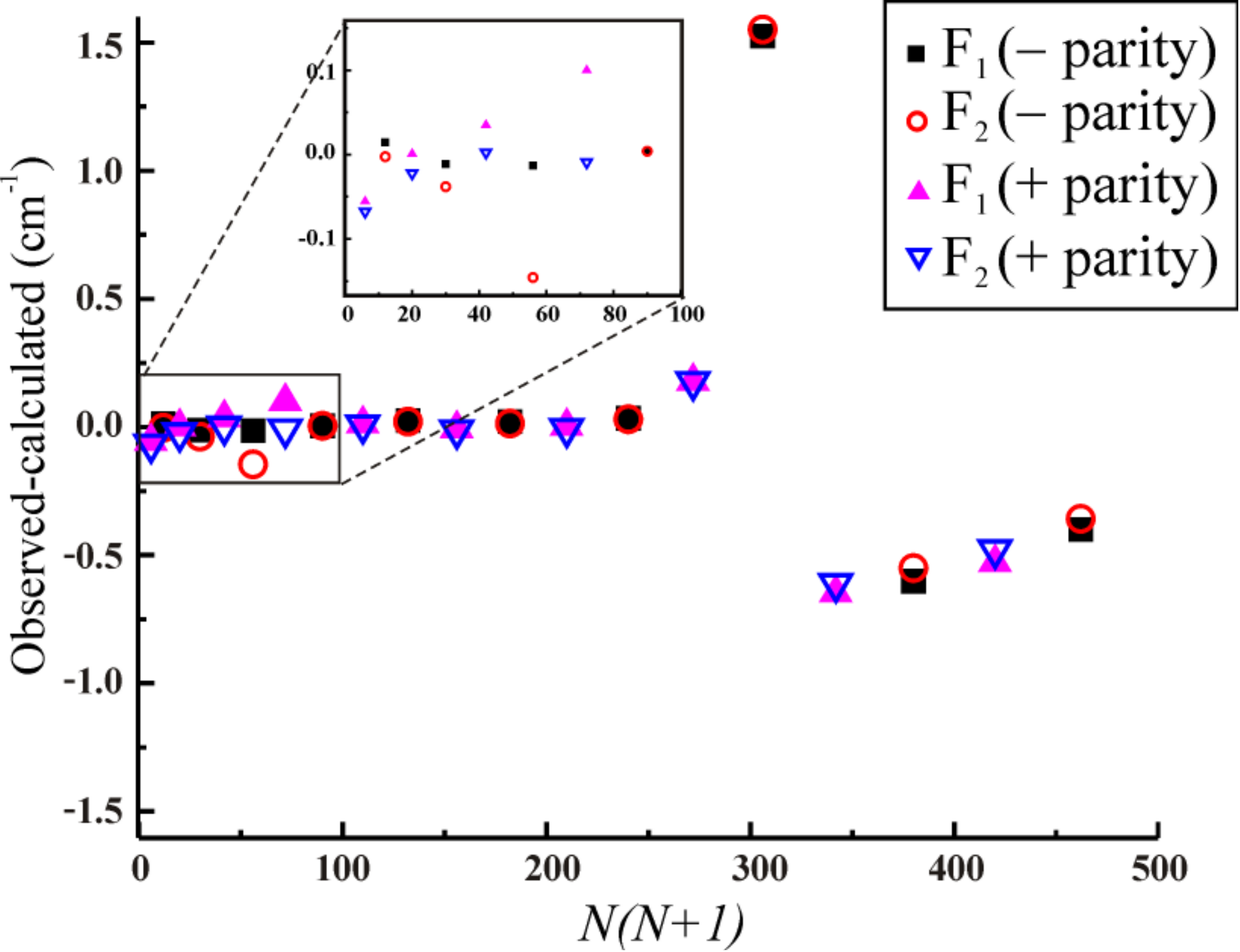}
		\centering
		\caption{\footnotesize{Plot of observed$-$calculated upper level term values illustrating the $N$-dependent perturbations in the  $X$(0,2$^{0}$,3) $A$(0,3,0)$^{0}_\kappa $ level ($\Sigma$ symmetry) at 7088 cm$^{-1}$. The square symbols indicate F$ _1$ ($ - $ parity); the hollow circle symbols are F$_2 $ ($ - $ parity); the triangle symbols are F$ _1 $ (+ parity); the hollow, upside down triangle symbols are F$ _2 $ (+ parity). The inset figure shows an enlargement of the region $N(N+1) <$ 100.}}
		\label{figure7.}
	\end{figure}

\begin{figure}[H!]
	\centering		
	\begin{tabular}{ll}
	  \includegraphics[scale=0.7]{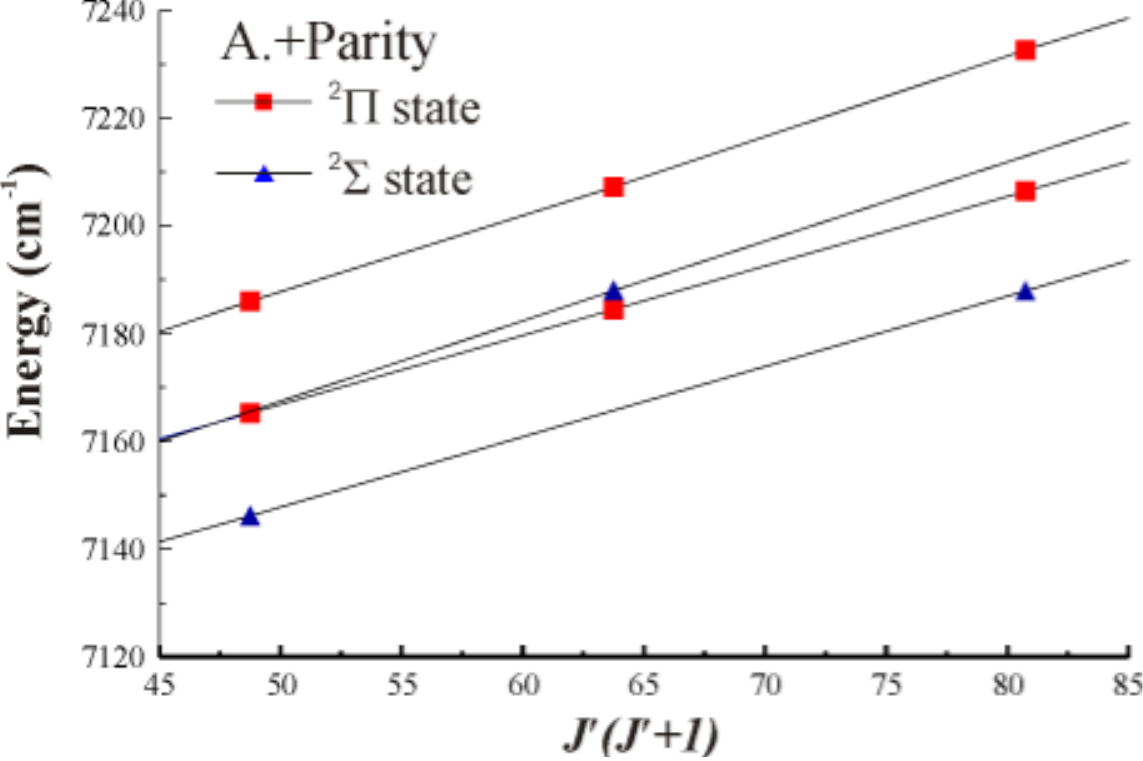}
	   &
	  \includegraphics[scale=0.7]{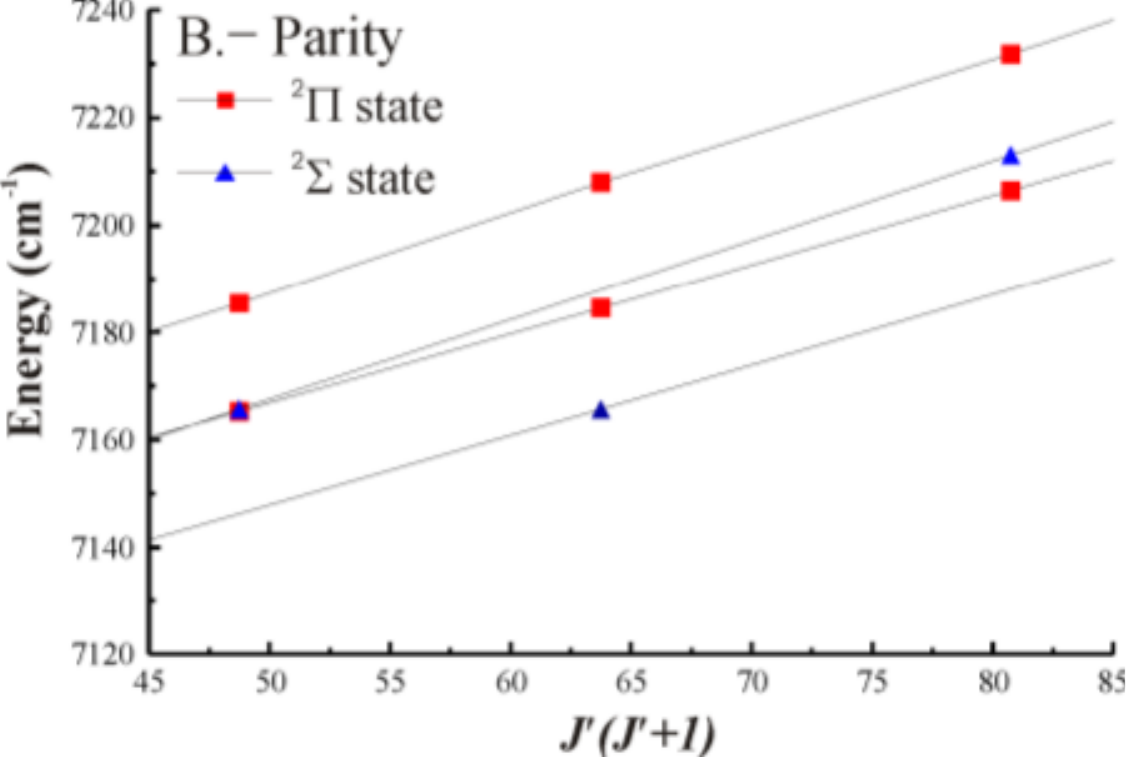}
        \end{tabular}	
		
	\caption{ \footnotesize{Parity-separated plots of the energy levels of the $^{2}\Sigma ^{+}$ state at 7088 cm$ ^{-1} $ and the $^{2}\Pi$ state at 7110 cm$ ^{-1} $  using parameters from Table \ref{table4}. }}
	\label{figure8.}
	\end{figure}

Neglecting small nuclear hyperfine coupling effects, the observed discrete $J$-dependent perturbations require perturbing levels of the same $J$ and parity to lie nearby in energy. By examining plots (Figure \ref{figure7.}) of observed$-$calculated energies against $N(N+1)$, for the upper $^2\Sigma ^+$ state of the band at 7088 cm$^{-1}$, local perturbations at $J^{\prime}=6.5$ ($-$ parity), $J^{\prime}=8.5$ (+ parity), and for all $J$ associated with $N^{\prime} \geq 15$ can be seen. For the first two levels, only one $J$-component of a given $N$ is perturbed, implying that the perturbing level must belong to a state where the parity does not\cite{Sharp-Williams2011a} alternate with $N$, \textit{i.e.} probably a nearby $^{2}\Pi $ state. However, the observed $^2 \Pi $ state at 7110 cm$^{-1}$ is \textit{not} the perturber because $J^{\prime}=6.5$ ($ - $ parity) of the $^2\Sigma ^+$ state is pushed down by a state that has $J^{\prime}=6.5$ at higher energy, while the $J^{\prime}=6.5$ ($-$ parity) of the 7110 cm$^{-1}$ $^2 \Pi $ state has a lower energy, by approximately 0.5 cm$^{-1}$, as shown in Figure \ref{figure8.}. The observed $^2\Pi$ state level in question is also perturbed and pushed down in energy, as can be seen in Table III.  Similarly, $J=8.5^+$ of the $^2\Sigma ^+$ state lies well below the energy of the corresponding level in the $^2\Pi$ state which therefore cannot cause the observed $^2\Sigma ^+$ perturbation either.  We conclude the perturbing level is a spectroscopically dark state.  Supplementary data available\cite{Tarroni2003} for the published calculations list all $^2\Sigma$ and $^2\Pi$ levels up to 10 000 cm$^{-1}$ in energy, but there is no other state which could plausibly lead to the low-J perturbations, implying it is actually a $^2\Delta$ or higher angular momentum state.   

Above $N=15$, the 7088 cm$^{-1}$ $^2\Sigma ^+$ state suffers a much larger perturbation as can be seen in Figure \ref{figure7.}.  This perturbation affects both $+$ and $-$ parity levels equally so is qualitatively different to that seen for the two lower rotational levels.  However, at these $J$-values, all C$_2$H states will be close to case(b), and so the arguments used to infer the character of the perturber above cannot be applied. The  $X^2\Sigma ^+\,(0,16^0,0)$ level is calculated\cite{Tarroni2003} to lie some 40 cm$^{-1}$ below the 7088 cm$^{-1}$ experimental level, and this could plausibly be the source of the observed high-$J$ perturbations in 7088 cm$^{-1}$ band.

 The $^2\Pi \leftarrow\, ^2\Sigma ^+$ band at 7110 cm$^{-1}$ shows strong $ R $ branches and much weaker \textit{P} branches. For example at the position of $P _{1}$(12.5) and  $ P_{2}$(11.5), the observed spectroscopic features are barely above the noise. This unusual intensity pattern is not unknown for vibronic transitions involving  $^2\Pi $ and $^2\Sigma ^+$ states\cite{Field2004} and occurs when $\Delta\Lambda= \pm$1 perturbations are present. The $ PQR $ branch intensity pattern for transitions terminating in the mixed level can be very complex. However, the $\Delta\Lambda= \pm$1 perturbation always affects the intensity of  $\Delta J$ =+1 ($ R$-branch) and $ - $1($ P$-branch) in the opposite direction.  If the perturbation is strong, a whole branch (e.g $ R$-branch) can be missing while the other (e.g. $ P$-branch) gets much higher intensity as was observed in NCO by Dixon \cite{Dixon1960}.
 
 The numerous additional lines observed in the prompt spectrum of the recorded region are likely hot bands of C$_2$H.  From the generally reliable excited vibronic energies calculated by Tarroni and Carter\cite{Tarroni2003} and experimental rotational constants for many low-lying excited levels, we may hope to identify some of these additional bands using combination differences, although we have so far been unsuccessful in this attempt. Clearly, the kinetic analysis of vibronic relaxation would be more compelling with secure spectral assignments.

\subsection{Energy Transfer and Reactivity}
Because many free radicals are formed with substantial internal energy, and undergo rapid reactions that can compete with relaxation, there is recent interest in improving theoretical treatments of non-thermal effects in combustion \cite{Miller2006,Burke2015} and atmospheric chemistry. \cite{Glowacki2012}  Energetic radicals or adducts may react completely, prior to the establishment of any local thermodynamic equilibrium of internal states, under conditions when the reaction rates and even branching ratios may be rapidly changing with internal energy.  The case of  C$ _{2} $H may be a good place to investigate the behavior at intermediate pressures of inert colliders, where the competition between relaxation and reaction can be tuned and observed, rather than deliberately suppressed. The vibronic mixing of the \textit{A} $^2\Pi$ and \textit{X} $^2\Sigma ^+$ states in  C$ _{2} $H can make the energy dependent reactivity more complex than in other radicals with a simpler electronic structure.

The time dependence of ground state absorption signals for a photolytically generated hot radical like C$_2$H will depend on the initial state distribution, the vibronic relaxation cascade due to energy transfer collisions with both inert and unsuccessfully reactive collision partners, as well as the reactive loss of the radicals in their evolving, nonthermal energy distribution. The energy dependence of the reaction rate plays a well-known and central role in the conventionally measured temperature-dependent rate constants of thermalized radicals. The energy transfer characteristics are typically less easily measured under the high energy conditions relevant to chemical activation or photodissociation. It is, however, frequently possible to devise conditions when the subtle details of the energy transfer are unimportant, and internal energy relaxation eigenvalues are well separated from the chemically significant eigenvalues of the master equation that describes the coupled reaction and relaxation processes. \cite{Barker2015,Miller2016}

Previous studies have shown dramatically increased reactivity of unrelaxed \cite{Tasaki1994} and partially relaxed \cite{Shokoohi1986}  C$_2$H radicals from the photodissociation of acetylene in collisions with H$_{2}$, for example. The thermal rate constant for this reaction near room temperature is 5.7 $\times$ 10$^{-13}$ cm$^{3}$ molec$^{-1}$ sec$^{-1}$, increasing  sharply with temperature, following nonlinear Arrhenius behavior.\cite{Peeters2002}   It may be important to distinguish between the temperature dependence of the thermal rate constant and the variable reactivity of internally excited states of the radical.  Theoretical calculations\cite{Wang1994, Wang2006} of possible vibrational mode-specific effects in this reaction and its reverse do not find vibrational enhancement of the reaction with stretching excitation of C$_2$H and even find an inhibition with bending excitation.  The potential energy surface\cite{Sharma2009} used for the full-dimensional dynamics calculations\cite{Wang2006} does not, however, include the low energy \textit{A} state, which has a barrierless path to H + C$_2$H$_2$, according to the calculations of Peeters et al.\cite{Peeters2002}  

The low energy \textit{A} $^2\Pi$ state of C$_2$H may contribute further complexity to an energy-dependent reactivity, as transition states and barriers to reactions may be quite different for \textit{A} and \textit{X} state interactions. The vibronic mixing of \textit{A} and \textit{X} states is strong enough, however, that they may not be kinetically distinct species, but rather an energy-dependent mixture, with a correspondingly mixed reactivity. \cite{Shokoohi1986} Even the lowest vibrationally excited states nominally belonging to the \textit{X} $^2\Sigma^+$ state of C$_2$H are calculated to have about 0.05 $A$-state character, and the zero-order vibrationless level of the $A$ state is quite evenly fragmented among five vibronic levels with energies between 3604 and 4593 cm$^{-1}$, none of which have more than 40\% $A$ state character.\cite{Tarroni2003}

In contrast to the reaction with hydrogen, the thermal rate constant for the reaction of C$_2$H with acetylene is much faster: 1.3 $\times$ 10$^{-10}$ cm$^3$ molec$^{-1}$ s$^{-1}$ near 300 K, with little temperature dependence, as reviewed by Van Look and Peeters.\cite{Peeters1995} The trends in transient ground state population illustrated in Figure \ref{figure5.} and even more so in Figure \ref{figure4.} for CF$_3$C$_2$H suggest that slowing down the relaxation of the hot radicals at lower buffer pressure inhibits the reactive loss. Such behavior can be expected for a barrierless association followed by a submerged barrier to fragmentation.\cite{Vu2013}

A comparison of the transient spectra of hot and cold band lines following C$_2$H$_2$ photodissociation similar to our Figure \ref{figure6.} was reported by Yan, et al.\cite{Yan1987b} This work has been widely cited \cite{Stephens1987,Lander1990,Farhat1993} as evidence that in the presence of ~20 Torr He, the vibrational thermalization of C$_2$H will be complete in less than a microsecond.    Without fully reproducing the conditions of Yan al. \cite{Yan1987b} we cannot be certain, but it appears likely that in the 3.5\% mixture of acetylene and He used for this measurement, the disappearance rate of vibrationally excited C$_2$H is dominated by reaction with C$_2$H$_2$ rather than vibrational relaxation by He. At the lower acetylene pressures typically used in subsequent kinetic experiments, the reactive loss is suppressed and vibrational excitation can be expected to persist to longer times.  SF$_6$ has often also been added to test gas mixtures to accelerate vibrational relaxation, although several workers have reported that it was found to have no effect on ground state kinetic spectroscopy of C$_2$H and was therefore left out.\cite{Stephens1987, Farhat1993, Opansky1993, Opansky1996} Until future experiments are able to provide a direct monitor of the vibrational relaxation, we can only echo a summary statement from Laufer and Fahr \cite{Laufer2004} in their 2004 review of the reactions and kinetics of ethynyl radicals from acetylene photodissociation:  ``the role of vibrational excitation in the subsequent chemistry is not clear."

As the computational machinery for \textit{ab initio} kinetics becomes more robust and well tested, the ambition grows to improve the treatment of chemical activation problems, including multiwell branching in competition with relaxation.\cite{Klippenstein2015}  Coupled electronic states add to the challenge.  Particularly for fast radical reactions that may react to completion faster than thermalization, a collection of thermal rate constants may not be sufficient to characterize a mechanism.  An extended set of state-resolved  C$_2$H kinetic measurements could provide a challenging target for the development of next-generation master-equation-based kinetic models.  The newly identified strong near-infrared bands can facilitate these measurements.

\section{Concluding remarks}
Three new near-infrared bands of C$_2$H have been observed and analyzed in the gas phase by diode laser spectroscopy. Transitions centered at 7110, 7088, and 6696 cm$^{-1}$ were found to be in good agreement with the calculated energies of vibronic levels of mixed $X^2\Sigma^+$ and $A^2\Pi$ electronic character.\cite{Tarroni2003}  The transitions are among the strongest of all those in the infrared and near infrared spectrum of the radical and provide a convenient spectroscopic tool for monitoring concentrations of the ground state radical in chemical mixtures.  Kinetic measurements show that the competition between reactive loss and vibrational relaxation of the initially hot radicals makes interpretation of the kinetic signals complicated, especially because of the presence of the low-lying excited electronic state which has different reactivity and whose wavefunction contaminates all vibrationally excited levels of the ground electronic state.  These attributes combine to make C$_2$H a particularly attractive vehicle for studying the effects of the presence of multiple potential surfaces on collisional 
processes in small free radicals where accurate calculations are now possible.   

\section{Supplemental material}
 See supplementary material for C$ _{2} $H spectral data in two regions between 6630 and 7135 cm$^{-1}$ obtained at 0, 2.6, and 3.6 $\mu $s time delay, using 1:1 mixtures of CF$_{3}$C$_{2}$H and Ar at a total pressure of 1.0 Torr.  Supplemental figure S1 and its caption, showing transient absorption signals for a selection of rotational levels in a 1\% mixture of CF$_3$C$_2$H in Ar at a total pressure of 5 Torr and a Boltzmann temperature analysis.

\section{Acknowledgments}
Work at Brookhaven National Laboratory was carried out under Contract No. DE-SC0012704 with the U.S. Department of Energy, Office of Science, and supported by its Division of Chemical Sciences, Geosciences and Biosciences within the Office of Basic Energy Sciences.  We thank Stephen Klippenstein for productive discussions and exploratory calculations on the submerged barriers in the C$_2$H + CF$_3$C$_2$H reaction.


%


\clearpage

\section{Figures and Tables}

\clearpage

	\clearpage

\begin{table}[f!]
	\centering
	\caption[Observed and calculated line positions in wavenumber (cm$^{-1}$) $^{2}\Sigma ^{+}\leftarrow\, ^{2}\Sigma ^{+}$(000) band of C$_{2}$H at 7087.645 cm$^{-1}$. The lines are marked with (*) are not included in the fit.]{Observed and calculated line positions in wavenumber (cm$^{-1}$) $^{2}\Sigma^{+}\leftarrow\, ^{2}\Sigma ^{+}$(000) band of C$_{2}$H at 7087.645 cm$^{-1}$. The lines are marked with (*) are not included in the fit.}
	\scriptsize
	\label{table1}
	\begin{tabular}{cdcdcdcdc}
		\hline
		\hline
		\ N$^{\prime\prime}$  & R$$_{1}$$  & obs$ - $calc &  R$$_{2}$$ & obs$ - $calc &  \ P$$_{1}$$ & obs$ - $calc & P$$_{2}$$ & obs$ - $calc \\ 
		
		\hline
		1	&	7093.035*	&	-0.055	&	7093.035*	&	-0.069	&			&		&			&		\\
		2	&	7095.640		&	0.016	&	7095.640		&	-0.003	&			&		&			&		\\
		3	&	7098.034		&	0.003	&	7098.034*	&	-0.022	&	7078.468*	&	-0.056	&	7078.468*	&	-0.066	\\
		4	&	7100.303		&	-0.009	&	7100.303*	&	-0.038	&	7075.245		&	0.013	&	7075.245		&	-0.002	\\
		5	&	7102.500*	&	0.035	&	7102.500		&	0.001	&	7071.812		&	-0.001	&	7071.812*	&	-0.022	\\
		6	&	7104.475		&	-0.014	&	7104.380*	&	-0.148	&	7068.256		&	-0.013	&	7068.256*	&	-0.038	\\
		7	&	7106.485*	&	0.102	&	7106.419		&	-0.009	&	7064.632*	&	0.035	&	7064.632		&	0.004	\\
		8	&	7108.153		&	0.006	&	7108.202		&	0.006	&	7060.786		&	-0.012	&	7060.691*	&	-0.143	\\
		9	&	7109.791		&	0.012	&	7109.843		&	0.010	&	7056.969*	&	0.098	&	7056.902		&	-0.009	\\
		10	&	7111.303*	&	0.026	&	7111.354*	&	0.018	&	7052.817		&	0.003	&	7052.862		&	0.002	\\
		11	&	7112.634		&	-0.007	&	7112.692		&	-0.013	&	7048.638		&	0.011	&	7048.684		&	0.007	\\
		12	&	7113.888*	&	0.020	&	7113.950		&	0.013	&	7044.333*	&	0.024	&	7044.387*	&	0.023	\\
		13	&	7114.955		&	-0.001	&	7115.023		&	-0.007	&	7039.850		&	-0.007	&	7039.907		&	-0.010	\\
		14	&	7115.939*	&	0.034	&	7116.014*	&	0.030	&	7035.293*	&	0.022	&	7035.348		&	0.012	\\
		15	&	7116.899*	&	0.188	&	7116.966*	&	0.171	&	7030.554		&	0.005	&	7030.614		&	-0.005	\\
		16	&	7118.899*	&	1.526	&	7118.987*$$^{a}$$	&	1.525	&	7025.726*	&	0.037	&	7025.791*	&	0.027	\\
		17	&	7117.238*	&	-0.650	&	7117.365*	&	-0.617	&	7020.877*	&	0.187	&	7020.936*	&	0.167	\\
		18	&	7117.652*	&	-0.602	&	7117.800*	&	-0.553	&	7017.073*	&	1.524	&	7017.208*	&	1.575	\\
		19	&	7117.937*	&	-0.531	&	7118.095*	&	-0.477	&	7009.619*	&	-0.645	&	7009.749*	&	-0.605	\\
		20	&	7118.129*	&	-0.400	&	7118.277*	&	-0.360	&	7004.234*	&	-0.599	&	7004.378*	&	-0.550	\\
		21	&			&		&			&		&	6998.728*	&	-0.526	&	6998.875*	&	-0.479	\\
		22	&			&		&			&		&	6993.122*	&	-0.402	&	6993.274*	&	-0.355	\\
						
		\hline
		
		\multicolumn{9}{c}{ $\sigma$=0.010 cm$^{-1}$}\\
		\hline
		\hline
		\multicolumn{9}{l}{$^{a}$Blended with \textit{R}$_{2}$(3) from $^{2}\Pi$ - $^{2}\Sigma ^{+}$(000) band of C$_{2}$H at 7109.646 cm$^{-1}$}\
	\end{tabular}
	\normalsize
\end{table}
\begin{table}[H!]
	\centering
	\caption[Observed and calculated line positions in wavenumber (cm$^{-1}$) $^{2}\Sigma ^{+} \leftarrow\, ^{2}\Sigma ^{+}$(000) band of C${2}$H at 6695.678 cm$^{-1}$. The lines are marked with (*) are not included in the fit.]{Observed and calculated line positions in wavenumber (cm$^{-1}$) $^{2}\Sigma ^{+}\leftarrow\, ^{2}\Sigma ^{+}$(000) band of C$_{2}$H at 6695.678 cm$^{-1}$. The lines are marked with (*) are not included in the fit.}
	\scriptsize
	\label{table2}
	\begin{tabular}{cdcdcdcdc}
	\hline
	\hline
	\ N$^{\prime\prime}$  & R$$_{1}$$ & obs$ - $calc &  R$$_{2}$$ & obs$ - $calc &  \ P$$_{1}$$ & obs$ - $calc & P$$_{2}$$ & obs$ - $calc \\ 
		
	\hline
	1	&	6701.285		&	-0.008	&	6701.285		&	-0.001	&			&		&			&		\\
	2	&	6703.991		&	-0.002	&	6703.991 	*	&	0.009	&	6689.773		&	-0.009	&	6689.773		&	-0.001	\\
	3	&	6706.622		&	0.003	&	6706.612		&	0.007	&	6686.717		&	-0.010	&	6686.717		&	0.001	\\
	4	&	6709.172		&	0.001	&	6709.151		&	-0.002	&	6683.600		&	-0.001	&	6683.589		&	0.003	\\
	5	&	6711.647		&	0.001	&	6711.626		&	0.001	&	6680.399		&	-0.002	&	6680.383		&	0.000	\\
	6	&	6714.045		&	0.001	&	6714.023		&	0.004	&	6677.133		&	0.005	&	6677.108		&	0.002	\\
	7	&	6716.365		&	0.004	&	6716.334		&	0.001	&	6673.786		&	0.007	&	6673.757		&	0.003	\\
	8	&	6718.596		&	0.000	&	6718.564		&	0.000	&	6670.355		&	0.002	&	6670.329		&	0.005	\\
	9	&	6720.744		&	-0.001	&	6720.705		&	-0.004	&	6666.856		&	0.007	&	6666.818		&	0.002	\\
	10	&	6722.805		&	-0.001	&	6722.757		&	-0.009	&	6663.265		&	0.002	&	6663.227		&	0.000	\\
	11	&	6724.724*	&	-0.050	&	6724.700*	&	-0.032	&	6659.594		&	0.001	&	6659.547		&	-0.006	\\
	12	&	6726.640		&	-0.008	&	6726.603		&	0.002	&	6655.835		&	-0.002	&	6655.788		&	-0.005	\\
	13	&	6728.426		&	0.004	&	6728.378		&	0.006	&	6651.934*	&	-0.057	&	6651.918*	&	-0.026	\\
	14	&	6730.117*	&	0.024	&	6730.064*	&	0.025	&	6648.043		&	-0.008	&	6647.997		&	-0.004	\\
	15	&	6731.704*	&	0.049	&	6731.652*	&	0.054	&	6644.019		&	0.004	&	6643.966		&	0.005	\\
	16	&	6733.200*	&	0.095	&	6733.147*	&	0.103	&	6639.898*	&	0.021	&	6639.845*	&	0.026	\\
	17	&			&		&			&		&	6635.679*	&	0.045	&	6635.626*	&	0.054	\\
	18	&			&		&			&		&	6631.375*	&	0.094	&	6631.315*	&	0.099	\\
		
		\hline

		\multicolumn{9}{c}{ $\sigma$=0.005 cm$^{-1}$}\\
\hline
\hline
		
	\end{tabular}
	\normalsize
\end{table}

\begin{table}[h!]
	\centering
	\caption[Observed and calculated line positions in wavenumber (cm$^{-1}$) $^{2}\Pi\leftarrow\,^{2}\Sigma ^{+}$(000) band of C$_{2}$H at 7109.646 cm$^{-1}$. The lines are marked with (*) are not included in the fit.]{Observed and calculated line positions in wavenumber (cm$^{-1}$) $^{2}\Pi\leftarrow\, ^{2}\Sigma ^{+}$(000) band of C$_{2}$H at 7109.646 cm$^{-1}$. The lines are marked with (*) are not included in the fit.}
	\scriptsize
	\label{table3}
	\begin{tabular}{c d c d c d c d c d c d c }
		\hline
		\hline
		\ N$^{\prime\prime}$  & R$$_{1}$$ & obs$ - $calc &  R$$_{2}$$ & obs$ - $calc &   Q$$_{1}$$ & obs$ - $calc &  Q$$_{2}$$ & obs$ - $calc &  P$$_{1}$$ & obs$ - $calc & P$$_{2}$$ & obs$ - $calc \\ 
		\hline
		2	&	&	&	&	&	7106.633	&	-0.003	&	7109.394		&	0.011	&	&	&	&	\\
		3	&	&	&	7118.985 *$$^{a}$$	&	-0.022	&	7106.287 *	&	-0.062	&	7108.538	&	-0.015	&	&	&	&		\\
		4	&	7119.585	&	-0.009	&	7120.878	&	-0.005	&	7105.685	*	&	-0.125	&	7107.783	*	&	0.087	&	7094.684	&	0.013	&	&		\\
		5	&	7121.538	*	&	-0.018	&	7122.627		&	-0.003	&	7105.060	&	0.009	&	7106.970	*	&	0.22	&	7091.194	*	&	-0.023	&	7092.767	*	&	-0.017	\\
		6	&	7123.318	&	-0.001	&	7124.239	&	0.010	&	7104.025 *	&	-0.063	&	7105.673	&	-0.009	&	7087.551	&	0.000	&	7088.835	&	0.000	\\
		7	&	7124.870		&	-0.013	&	7125.672	&	0.001	&	7102.983 *	&	0.040	&	7104.474	&	-0.002	&	7083.682	&	-0.007	&	7084.759	&	0.001	\\
		8	&	7126.259	&	-0.001	&	7126.952	&	0.004	&	7101.729	*	&	0.103	&	7103.134		&	0.013	&	7079.623	&	-0.006	&	7080.539	&	0.005	\\
		9	&	7127.446	&	-0.004	&	7128.061	&	0.005	&	&	&	&	&	&	&	7076.159	&	0.005	\\
		10	&	7128.450		&	-0.006	&	7128.981		&	-0.010	&	&	&	7099.934	&	-0.006	&	7070.934		&	0.007	&	7071.617	&	0.006	\\
		13	&	7129.277	&	-0.001	&	7130.755	&	0.006	&	7096.621	&	0.007	&	7098.101	&	-0.006	&	7066.297	&	-0.002	&	7066.903	&	0.003	\\
		11	&	7129.906	&	-0.008	&	7129.748	&	-0.004	&	7094.623	&	0.030	&	&	&	7061.482	&	-0.006	&	7062.016	&	-0.003	\\
		12	&	7130.378	&	0.012	&	7130.318	&	-0.017	&		&	&	7093.971	&	0.028	&	7056.497	&	0.003	&		&		\\
		14	&	7130.642	&	0.011	&	7130.961	&	0.001	&	7090.024	&	-0.002	&	&	&	7051.313	&	-0.005	&	7051.720		&	-0.014	\\
		15	&	7130.700	&	-0.009	&	7131.016	&	0.016	&	7087.466	&	-0.013	&	7089.087	&	-0.015	&	7045.971	&	0.013	&	7046.324	&	-0.014	\\
		16	&		&		&	7130.906	*	&	0.052	&	&		&	&	&	&	&	&		\\
	\hline

		\multicolumn{13}{c}{$\sigma$=0.011 cm$^{-1}$}\\
		\hline
		\hline
		\multicolumn{9}{l}{$^{a}$Blended with \textit{R}$_{2}$(16) from $^{2}\Sigma^{+}$ - $^{2}\Sigma ^{+}$(000) band of C$_{2}$H at 7087.645 cm$^{-1}$}\	
	\end{tabular}
	\normalsize
\end{table}

\begin{table}[h]
	\centering
	\caption[Spectroscopic parameters in wavenumber (cm$^{-1}$) of observed C$_2$H bands]{Spectroscopic parameters in wavenumber (cm$^{-1}$) of observed C$_2$H bands}
	\scriptsize
	\label{table4}
	\begin{tabular}{ccccc}
		\hline
		\hline
		\   & X(0,0,0) &   X(0,8$^{0}$,2) & X(0,2$^{0}$,3)A(0,3,0)$^{0}_\kappa$& X(0,3$^{1}$,3)A(0,0,2)$^{1}$ \\
		\   &$^{2}\Sigma ^{+ a}$	&$^{2}\Sigma ^{+}$ & $^{2}\Sigma ^{+}$& $^{2}\Pi $\\
		\hline
		A & & & & -4.29(2) \\
		B & 1.4568256(2) & 1.42115(5)       & 1.3941(2) & 1.3719(1)\\
		Dx10$^{5}$ & 0.357(3) & 1.582(3) & 0.89(7) & 0.68(4) \\
		$ \gamma $& 0.0020883(1) & 0.0015(2)&  -0.0071(5)   & -0.0336(6) \\
		p+2q & & & & -0.0925(8) \\
		q & & & & -0.00447(3)\\
		Origin & & 6695.678(2) & 7087.645(7) & 7109.646(6)\\
		\hline
	    &\multicolumn{4}{c}{Calculated values$ ^{b} $}\\
		\hline
		Origin & & 6696.7 & 7099.0  & 7106.7 \\
		Intensity & &	327.5 & 1384.1  &  744.2\\		
		\hline
			\hline
		\multicolumn{5}{l}{a. The ground state constants from Gottlieb et al.\cite{Gottlieb1983}}\\
		\multicolumn{5}{l}{b. Predicted values are from Tarroni and Carter \cite{Tarroni2004}} \\
		
	\end{tabular}
	\normalsize
\end{table}

\end{document}